\documentclass[usenatbib,letters]{mnras}
\usepackage{txfonts}
\usepackage{graphicx}
\usepackage[Symbol]{upgreek}
\usepackage[T1]{fontenc}
\usepackage[utf8]{inputenc}

\title[The centre of GRS\,J0844$+$4627 radio galaxy]{The structure at the 
centre of the giant radio galaxy GRS\,J0844$+$4627:\\a compact symmetric 
object?}
\author[A. Marecki, B. Sebastian, C. H. Ishwara-Chandra]
{A. Marecki$^{1}$\thanks{E-mail: \texttt{amr@astro.uni.torun.pl}}, B. Sebastian$^{2}$, C. H. Ishwara-Chandra$^{3}$\\
$^{1}$Institute of Astronomy, Nicolaus Copernicus University, Faculty of Physics, Astronomy and Informatics, ul. Grudziądzka 5,
PL-87-100 Toruń, Poland\\
$^{2}$Department of Physics and Astronomy, University of Manitoba, Winnipeg, Canada\\
$^{3}$National Centre for Radio Astrophysics, TIFR, Post Bag No.\,3, Ganeshkhind Post, 411007 Pune, India}

\date{Accepted 2022 September 01. Received 2022 August 29; in original form 2022 June 05}

\pubyear{2022}

\begin{document}
\label{firstpage}
\pagerange{\pageref{firstpage}--\pageref{lastpage}}
\maketitle

\begin{abstract}

We observed the core region of the giant radio galaxy 
GRS\,J0844$+$4627 with e-MERLIN at 1.52 and 5.07\,GHz. These observations 
revealed that the apparent single feature at the centre of GRS\,J0844$+$4627, 
as seen by GMRT, consists of two components separated by 2.7\,kpc in 
projection. Follow-up observations at 1.66\,GHz using the EVN 
unveiled the complex 
morphologies of the two components. In particular, the south-western 
component identified with the SDSS\,J084408.85+462744.2 galaxy
morphologically resembles a compact symmetric object (CSO) with a 
projected linear size of 115\,pc. If the CSO hypothesis turns out to 
be correct, then the overall radio structure of GRS\,J0844$+$4627 is 
triple--double. Given that CSOs are considered young objects, 
GRS\,J0844$+$4627 would appear as a recently restarted active galaxy.

\end{abstract}


\section{Introduction}

GRS\,J0844$+$4627 radio source was observed with GMRT\footnote{Giant Metrewave 
Radio Telescope \citep{Swarup1991}} at 150\,MHz when searching the 
Leiden--Berkeley Deep Survey (LBDS) Lynx field for high-redshift radio galaxies 
\citep{Ishwar2010} and followed up with GMRT at 325, 610, and 
1250\,MHz \citep[][hereafter Paper\,I]{Sebastian2018}. Based on all four 
observations mentioned above, it has been found that the central region of 
GRS\,J0844$+$4627, resolved by neither the GMRT at any frequency nor by the VLA 
at 1.4\,GHz (FIRST\footnote{Faint Images of the Radio Sky at 
Twenty-Centimeters\\ \citep{FIRST}} survey), has a~steep spectrum: 
$\upalpha\!=\!-0.85\pm0.032$ ($S\!\propto\!\upnu^{\upalpha}$) -- see fig.\,6 in 
Paper\,I -- thus, it belongs to a~class of compact steep-spectrum (CSS) 
sources. It is identified with SDSS\,J084408.85+462744.2 galaxy at the 
redshift of $z\!=\!0.5696228$. For this redshift and the cosmological 
parameters specified at the end of this section, the radio galaxy's overall 
angular size of 5.5\,arcmin translates to 2.2\,Mpc of linear size hence
the object qualifies as a giant radio galaxy.

In some sense, GRS\,J0844$+$4627 resembles TXS\,0818$+$214. The latter radio 
source appears in FIRST as a triple, where the central component also has a 
steep spectrum. Owing to the high resolution provided by
MERLIN\footnote{Multi-Element Remotely-Linked Interferometer Network\\
(\texttt{www.e-merlin.ac.uk})} and the
EVN\footnote{European VLBI Network (\texttt{www.evlbi.org})} observations,
\citet{Marecki2009} unveiled the nature of the apparent core of TXS\,0818$+$214
-- it is a double. Consequently, TXS\,0818$+$214 turned out to be a
double--double radio source \citep{Schoenmakers2000}. As a follow-up of GMRT 
observations (Paper\,I), we carried out an observing programme similar to 
that by \citet{Marecki2009} in order to study the CSS
source at the centre of GRS\,J0844$+$4627. The outcome thereof is 
reported here.

For consistency with Paper\,I, the cosmological parameters used here are: 
$H_0\!=\!69.3\,{\rm km\,s}^{-1}{\rm Mpc}^{-1}, \omega_{\rm M}\!=\!0.286$, 
and $\omega_{\rm baryon}\!=\!0.0463$. One arcsec of angular size translates to 
6.625\,kpc of linear size for these values.

\section{Observations and their results}

\subsection{e-MERLIN observations}

We observed the CSS source at the centre of GRS\,J0844$+$4627 with e-MERLIN 
at 1.52\,GHz (512-MHz bandwidth) on 2018~August~31 and at 5.07\,GHz 
(512-MHz bandwidth) on 2019~September~14. 
Only Mark2, Knockin, Defford, and Cambridge antennas observed at 
1.52\,GHz hence the shortest baselines were missing. The observing time was 
extended to 72\,h to compensate for the sensitivity loss caused by the 
missing antennas. Mark2, Defford, Pickmere, Darnhall, and Cambridge 
observed at 5.07\,GHz, whereas Knockin was missing thus some of the long 
baselines were absent at 5.07\,GHz. The duration of that observation was 
20\,h. The very different distribution of the spatial frequencies led to a 
similar restoring beam size for the 1.52- and 5.07-GHz data. The data reduction
was carried out in {\sc aips}.\footnote{Astronomical Image Processing System
(\texttt{www.aips.nrao.edu})}

\begin{figure*}
\includegraphics[width=0.45\linewidth]{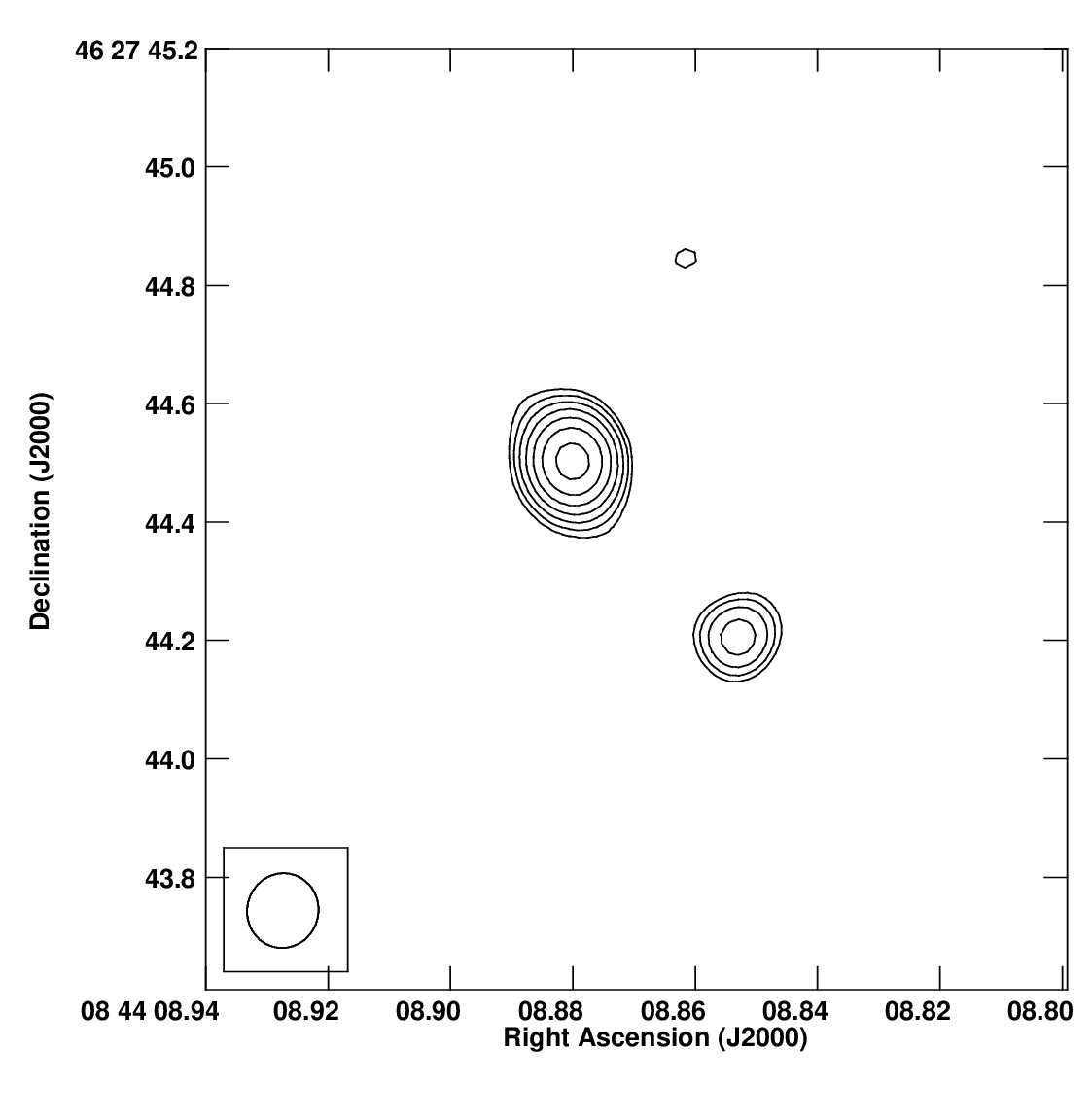}
\includegraphics[width=0.45\linewidth]{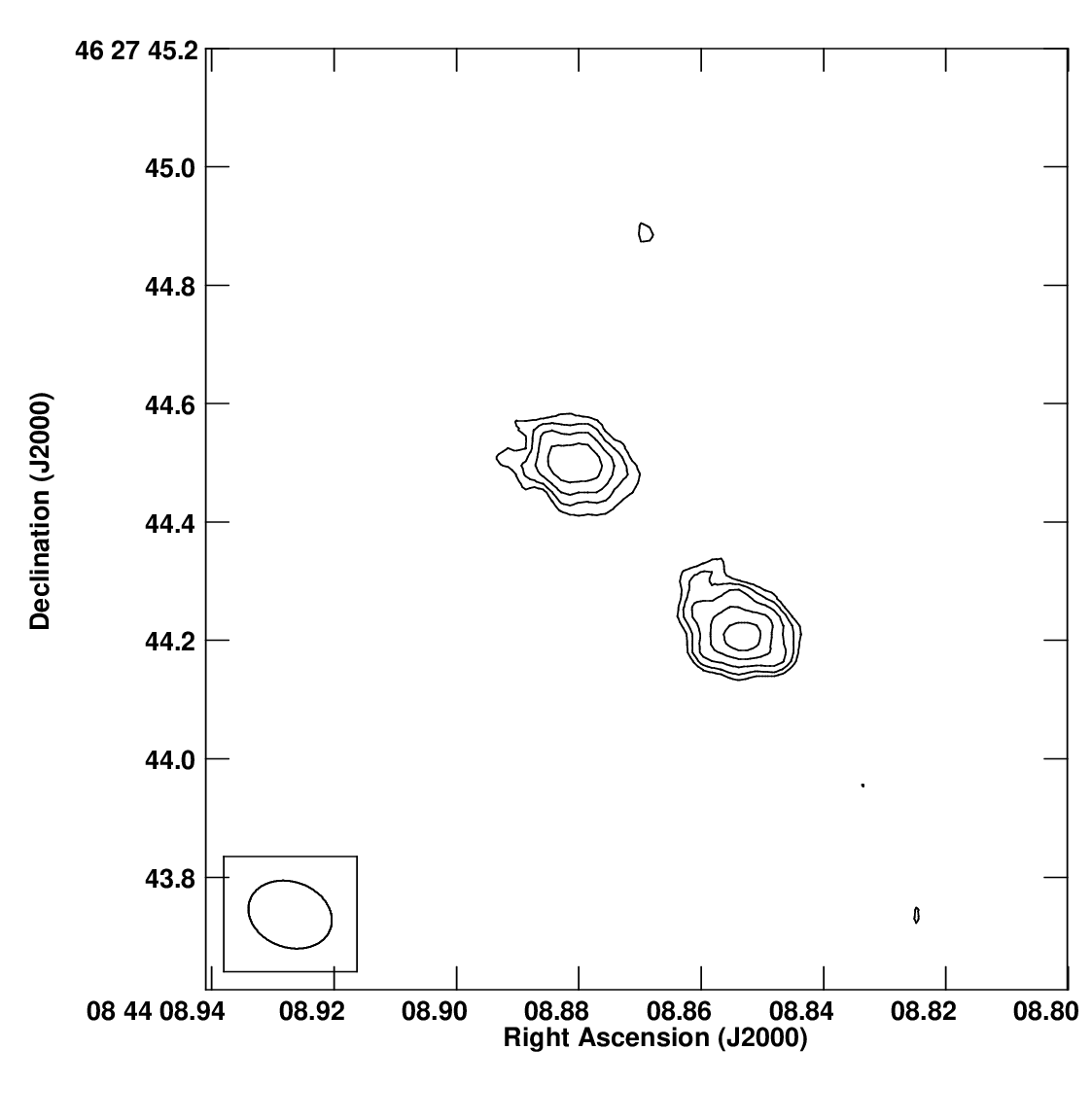}
\caption{e-MERLIN images of the central region of GRS\,J0844$+$4627 at 
1.52\,GHz (left-hand panel) and 5.07\,GHz (right-hand panel). Contours increase 
by a factor of $\sqrt2$; the first contour level is $70\,\upmu$Jy per beam 
for the 1.52-GHz image and $52\,\upmu$Jy per beam for the 5.07-GHz image. They both 
correspond to a $4\upsigma$ level. The beam sizes are 
$127\times120$\,mas at the position angle of $-13\degr$ for the 
1.52-GHz image and $144\times110$\,mas at the position angle of 
$70\degr$ for the 5.07-GHz image.}
\label{fig:MERLIN}
\end{figure*}

\begin{table}
\caption{Flux densities and spectral indices of the two features at the 
centre of GRS\,J0844$+$4627 as measured with e-MERLIN.}
\begin{center}
\begin{tabular}{c c c r}
\hline
Component & \multicolumn{2}{c} {Flux density [$\upmu$Jy]} & \multicolumn{1}{c}{$\upalpha$} \\
& 1.52\,GHz & 5.07\,GHz \\
\hline
NE & 693$\pm34$ & 222$\pm25$ & $-0.94$ \\
SW & 188$\pm27$ & 216$\pm23$ & 0.12 \\
\hline
\end{tabular}
\end{center}
\label{table:spix}
\end{table}

The images resulting from our 1.52- and 5.07-GHz e-MERLIN observations are 
shown in Fig.\,\ref{fig:MERLIN}. The apparent single CSS source at 
the centre of GRS\,J0844$+$4627, as seen by GMRT, reveals two 
components separated by 0.4\,arcsec (2.7\,kpc in projection). The line connecting them 
has a position angle (PA) of $43\degr$, coinciding with the PA of the 
large-scale lobes (Paper\,I, fig.\,2). The optical object 
SDSS\,J084408.85+462744.2 is identified with the south-western component. We 
measured the flux densities of those two components at both frequencies and 
calculated the spectral indices -- see Table\,\ref{table:spix}. The 
uncertainties on the flux densities have been estimated using {\sc aips} 
task \texttt{JMFIT}. The sum of the 1.52-GHz flux densities given in 
Table\,\ref{table:spix} is consistent with the flux density listed in FIRST 
(950\,$\upmu$Jy). The missing flux may be due to the poor sensitivity of 
e-MERLIN to possible extended structures caused by the absence of the two 
telescopes forming the three shortest baselines. Both components remain 
unresolved at 1.52\,GHz and only partly resolved at 5.07\,GHz when observed 
with e-MERLIN. To reveal the details of their morphology, we performed an EVN 
follow-up observation.

\begin{figure}
\includegraphics[width=1.0\linewidth]{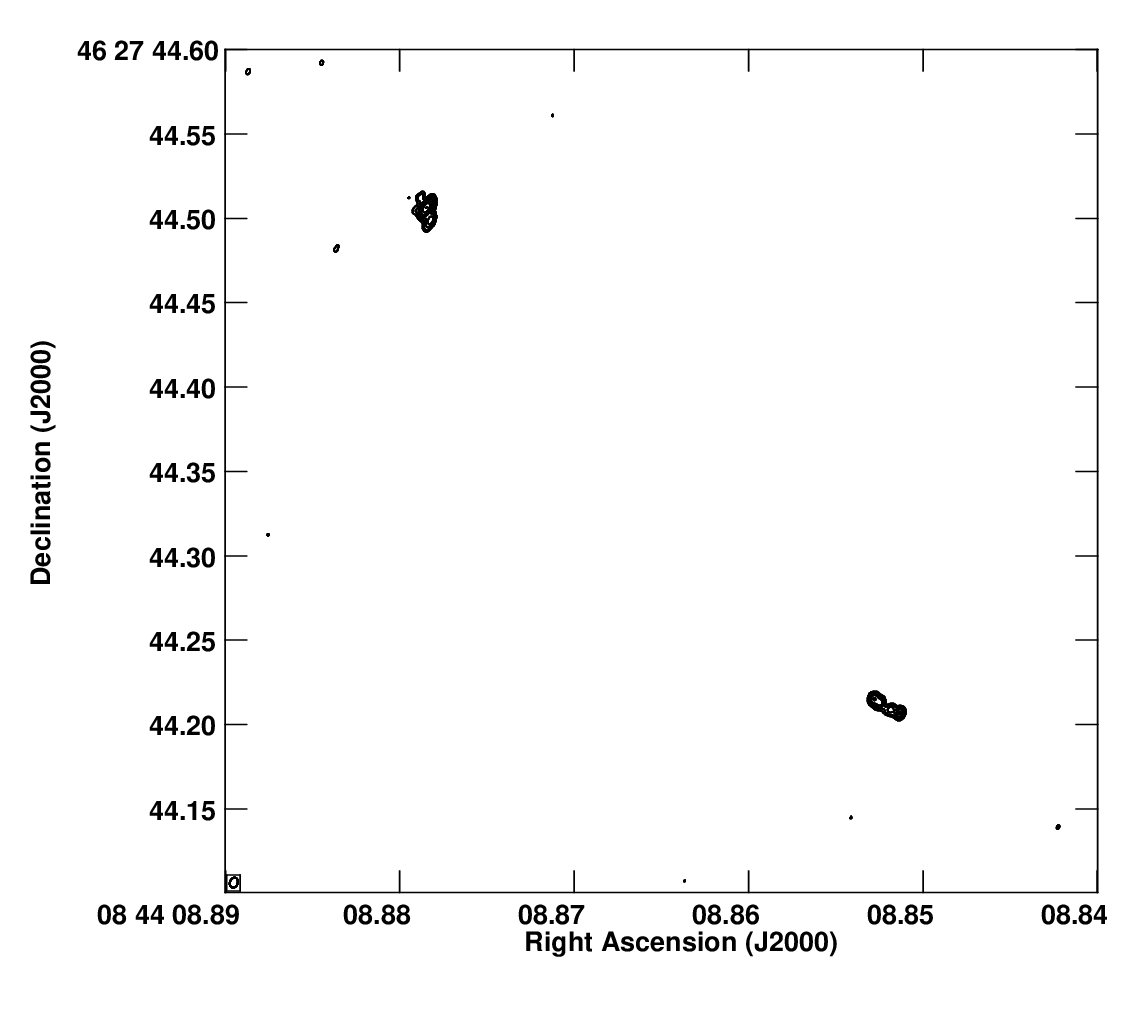}
\caption{EVN image of the central region of GRS\,J0844$+$4627 at 1.66\,GHz.
Contours increase by a factor of $\sqrt2$; the first contour level 
is $15.8\,\upmu$Jy per beam, which corresponds to a $5\upsigma$ level. The 
beam size is $6.08\times 4.59$\,mas at the position angle of $-20\degr$.}
\label{fig:EVN_both}
\end{figure}

\begin{figure*}
\includegraphics[width=0.45\linewidth]{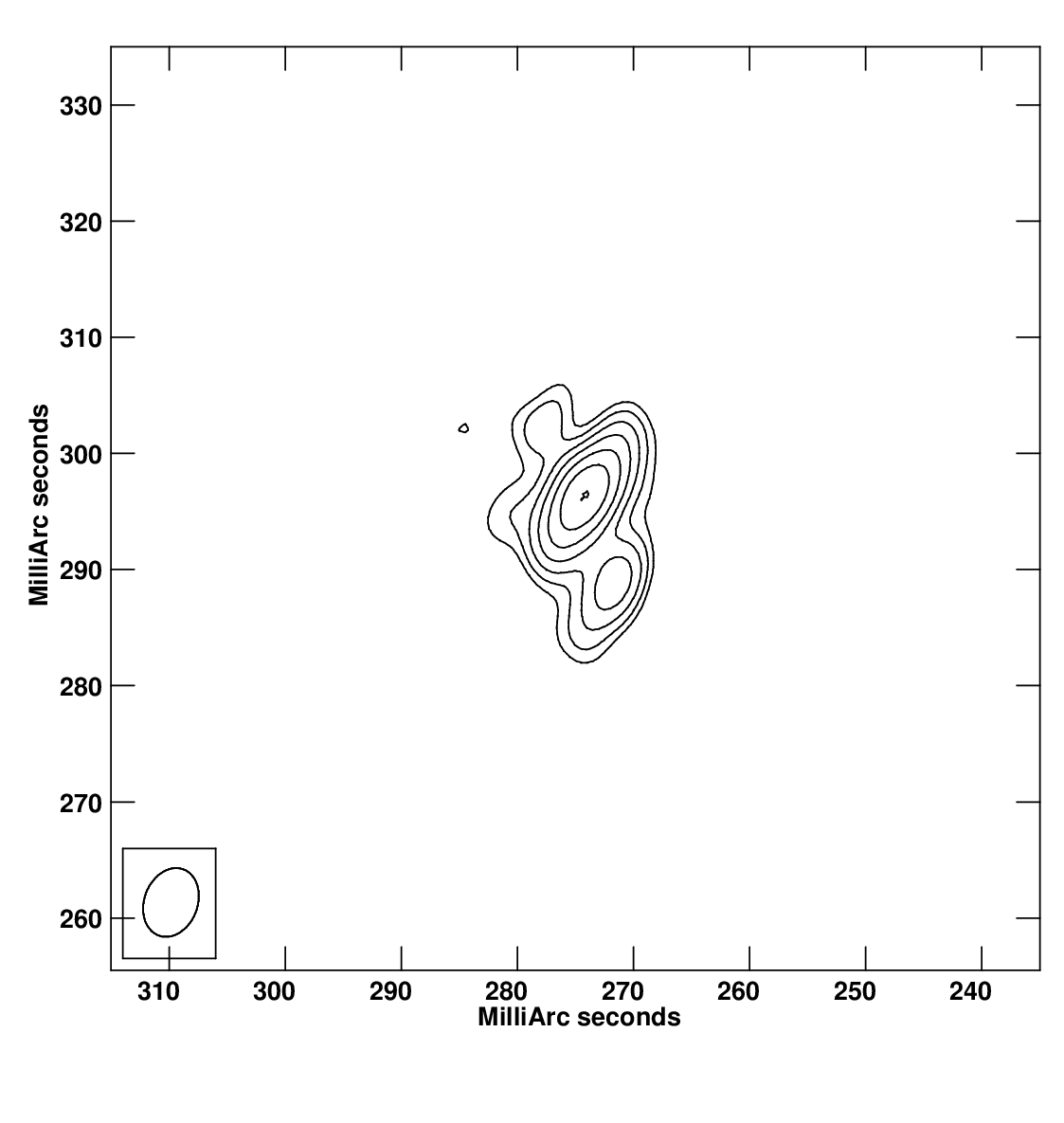}
\includegraphics[width=0.45\linewidth]{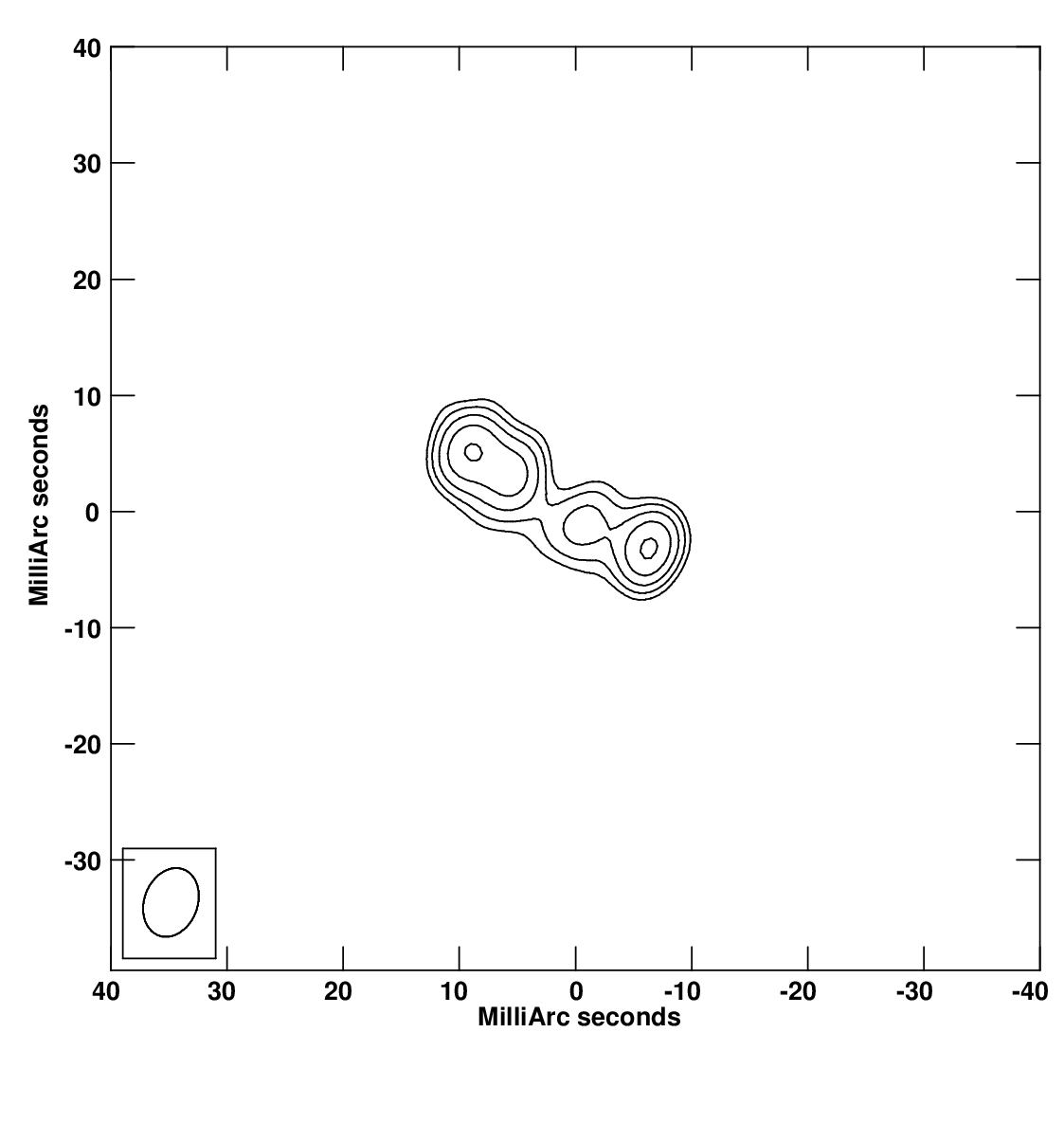}
\caption{Enlarged north-eastern (left-hand panel) and south-western
(right-hand panel) components of the EVN image of GRS\,J0844$+$4627
shown in Fig.\,\ref{fig:EVN_both}. The relative coordinates of both 
images are centred at RA\,$=\!08^{\rm h} 44^{\rm m} 08\fs8519331$,
Dec.\,$=\!+46\degr 27\arcmin 44\farcs210000$.
Contours increase by a factor of $\sqrt2$; the first contour level 
is $15.8\,\upmu$Jy per beam, which corresponds to a $5\upsigma$ level. The 
beam size is $6.08\times 4.59$\,mas at the position angle of $-20\degr$.}
\label{fig:EVN_enlarged}
\end{figure*}

\begin{table*}
\caption{Flux densities of the three subcomponents of the south-western 
component as measured with the EVN at 1.66\,GHz.}
\begin{tabular}{l c c r r c c c c c}
\hline
Subcomponent & Flux density & & \multicolumn{3}{c}{Size} & & \multicolumn{3}{c}{Deconvolved size}\\
             &              & & Major axis & Minor axis & PA & & Major axis & Minor axis & PA\\
             & [$\upmu$Jy]    & & \multicolumn{2}{c} {[mas]} & [$\degr$] & & \multicolumn{2}{c} {[mas]} & [$\degr$]\\
\hline
Eastern      & $84\pm4$     & & $6.57\pm0.18$  &  $4.59\pm0.13$ & $-20\pm3$ & & 3.10 & 1.25 & 10 \\
Central      & $41\pm2$     & & $6.50\pm0.34$  &  $4.59\pm0.24$ & $-20\pm6$ & & 3.47 & 1.73 & 18 \\
Western      & $68\pm2$     & & $6.14\pm0.19$  &  $4.59\pm0.15$ & $-20\pm4$ & & 2.19 & 1.33 & 21 \\
\hline
\end{tabular}
\label{table:SW_subcomp}
\end{table*}

\subsection{The EVN observation}

The 12-hour EVN observation of GRS\,J0844$+$4627 (project EM146)
was conducted on 2021~May~28 at 1.66\,GHz using 17\,radio telescopes, including
Lovell and four e-MERLIN antennas. JVAS\,J0847$+$4609 was used as a phase 
calibrator. The length of the target-calibrator cycle was 10\,min, of which 
7\,min~40\,s was spent on the target source. The data were recorded at
1-Gbps rate in four 32-MHz-wide sub-bands per each of the two circular 
polarizations. For e-MERLIN stations, the recording rate was 
512\,Mbps and there were two sub-bands per polarization. For 
these parameters, the image thermal noise estimated using the EVN
Calculator\footnote{\texttt{http://old.evlbi.org/cgi-bin/EVNcalc.pl}}
amounts to $3.16\,\upmu$Jy per beam ($1\upsigma$). The data were correlated at 
JIVE\footnote{Joint Institute for VLBI ERIC (\texttt{www.jive.eu})}. We 
carried out the reduction of the resulting visibilities in {\sc 
aips} in a~standard way.

The EVN image of the central region of GRS\,J0844$+$4627 is shown in 
Fig.\,\ref{fig:EVN_both}. As in Fig.\,\ref{fig:MERLIN}, two components 
are present; here, they are separated by 0.409\,arcsec (2.71\,kpc in projection) at the 
${\rm PA}\!=\!42\fdg9$. To make it possible to explore each of 
them in detail, they have been featured separately in Fig.\,\ref{fig:EVN_enlarged}. 
The south-western component (Fig.\,\ref{fig:EVN_enlarged}, right-hand panel) is a 
triple of 17.4\,mas angular size, i.e. 115\,pc of the projected linear size.
Its axis is oriented at the ${\rm PA}\!=\!62\fdg7$. The flux 
densities, sizes, and deconvolved sizes of the three subcomponents of the SW 
component fitted with \texttt{JMFIT} are shown in Table\,\ref{table:SW_subcomp}.
Since the north-eastern component is diffuse, fitting Gaussian subcomponents 
with \texttt{JMFIT} was problematic. The flux density of that region estimated
with the {\sc aips} \texttt{TVSTAT} utility amounts to $\sim0.4$\,mJy.

\section{Interpretation of the observations}

Based on the morphology of the SW component, it could be 
hypothesized that it is a compact symmetric object 
\citep[CSO,][]{Wilkinson1994}. CSOs are objects with sub-kpc linear sizes 
and a two-sided structure expanding at rates exceeding 0.1\,$c$. Both these 
factors imply low kinematic age; in the case of GRS\,J0844$+$4627 it is of 
the order of $10^3$\,yr. On the other hand, the spectral age of the 
large-scale structure of GRS\,J0844$+$4627 may reach 20\,Myr (Paper\,I), so 
if the CSO hypothesis is correct, then GRS\,J0844$+$4627 manifests 
restarted activity and the CSO is the emanation of the most recent active 
period. Large-scale radio emission has been found around several compact 
radio sources -- see section\,3.3 of \citet{ODea2021} -- which remains in 
line with the notion that the compact triple at the position of 
SDSS\,J084408.85+462744.2 is a~rejuvenated radio source -- see table\,1 in 
\citet{ODea2021}. By taking the size as an indicator of kinematic age, this 
looks like one of the youngest and smallest known CSOs.

It has been claimed in Paper\,I that GRS\,J0844$+$4627 is a double--double
source. Due to the presence of the putative CSO at the centre 
of the source, the overall radio structure appears to be triple--double.
Despite the number of active episodes, the orientation of each pair of 
lobes has not changed significantly, unlike in Seyfert galaxies, where the 
elements pertinent to the different active episodes are often misaligned 
with each other \citep[see e.g.][]{Sebastian2019}. The consistency in the 
direction could be because, unlike the smaller black holes in Seyferts, the 
larger-mass black holes hosted by elliptical galaxies are more resistant to 
spin flips.

The nature of the north-eastern component (Fig.\,\ref{fig:EVN_enlarged}, left-hand panel) 
is unclear. Its highly steep spectrum and the lack of the optical identification
suggest that it could be a relic lobe, but the putative south-western 
counterpart is neither visible in the e-MERLIN images nor the EVN ones.
There is a possibility that the 
NE component is a knot due to local inhomogeneity and no such knot 
has been formed in the SW direction. The spectral index of the NE component might 
have also been affected by resolution, i.e. there could be missing-flux 
issues at higher frequencies driving the spectral index to a steeper side, 
though knots are not expected to have steep spectral indices.

Confirming the CSO hypothesis would require measuring the 
subcomponents' motion or establishing their spectral indices. In either 
case, new observations are necessary. If the CSO hypo\-thesis turns out to be
false, then the SW component is a core-jet source. This interpretation could 
explain the NE component as a part of the jet further away. Under the 
assumption that the SW component has a core-jet structure, the 
jet/counter-jet ratio is $\gtrsim$18 based on the flux densities from our 
image. While the inclination angle and the bulk velocities both determine 
the jet/counter-jet ratio, we make some assumptions about the inclination 
angle since GRS\,J0844$+$4627 is a giant radio galaxy. The size of the
largest known giant radio galaxy is $\sim5$\,Mpc \citep{Oei2022}. If the 
deprojected size of GRS\,J0844$+$4627 is also 5\,Mpc then the inclination 
angle would be $26\degr$. For a typical spectral index of $\upalpha\sim-0.6$ 
\citep{LB2013}, the bulk velocity turns out to be 0.56\,$c$. Given that such 
velocities are not unusual in core-dominated quasars \citep{ODea1988}, the 
core-jet scenario cannot be ruled out at the present stage.

\section*{Acknowledgements}

MERLIN is a National Facility operated by the University of Manchester at 
Jodrell Bank Observatory on behalf of STFC.\\
This work has benefited from research funding from the European Community's 
sixth Framework Programme under RadioNet R113CT 2003 5058187.\\
We thank Thomas Muxlow and Javier Moldon from Jodrell Bank Centre 
for Astrophysics for their help with e-MERLIN data reduction.\\
The European VLBI Network is a joint facility of independent European, African,
Asian, and North American radio astronomy institutes. Scientific results from
data presented in this publication are derived from the EVN project EM146.\\
This research has made use of the NASA/IPAC Extragalactic Database (NED),
which is funded by the National Aeronautics and Space Administration and
operated by the California Institute of Technology.\\
CHIC acknowledges the support of the Department of Atomic Energy,
Government of India, under the project 12-R\&D-TFR-5.02-0700.

\section*{Data availability}

The data underlying this paper will be shared on reasonable request to the 
corresponding author.

%%%%%%%%%%%%%%%%%%%%%%%%%%%%%%%%%%%%%%%%%%%%%%%%%%%%%%%%%%%%%%%%%%%%%%%%%%%%%%%%%%%%%%%%%%%%%%%%%%%%%%%%%%%%%%%%%%%%%%%%%%%%%%%%%%

\label{lastpage}


\begin{thebibliography}{}

\bibitem[\protect\citeauthoryear{Becker, White, \& Helfand}{1995}]{FIRST}
Becker R.~H., White R.~L., Helfand D.~J., 1995, ApJ, 450, 559

\bibitem[\protect\citeauthoryear{Ishwara-Chandra et al.}{2010}]{Ishwar2010}
Ishwara-Chandra C.~H., Sirothia S.~K., Wadadekar Y., Pal S., Windhorst R., 2010, MNRAS, 405, 436

\bibitem[\protect\citeauthoryear{Laing \& Bridle}{2013}]{LB2013}
Laing R.~A., Bridle A.~H., 2013, MNRAS, 432, 1114

\bibitem[\protect\citeauthoryear{Marecki \& Szablewski}{2009}]{Marecki2009}
Marecki A., Szablewski M., 2009, A\&A, 506, L33

\bibitem[\protect\citeauthoryear{O'Dea, Barvainis, \& Challis}{1988}]{ODea1988}
O'Dea C.~P., Barvainis R., Challis P.~M., 1988, AJ, 96, 435

\bibitem[\protect\citeauthoryear{O'Dea \& Saikia}{2021}]{ODea2021}
O'Dea C.~P., Saikia D.~J., 2021, A\&ARv, 29, 3

\bibitem[\protect\citeauthoryear{Oei et al.}{2022}]{Oei2022} 
Oei M.~S.~S.~L., et al., 2022, A\&A, 660, A2

\bibitem[\protect\citeauthoryear{Schoenmakers et al.}{2000}]{Schoenmakers2000}
Schoenmakers A.~P., de Bruyn A.~G., R{\"o}ttgering H.~J.~A., van der Laan H., Kaiser C.~R., 2000, MNRAS, 315, 371

\bibitem[\protect\citeauthoryear{Sebastian et al.}{2018}]{Sebastian2018}
Sebastian B., Ishwara-Chandra C.~H., Joshi R., Wadadekar Y., 2018, \mbox{MNRAS}, 473, 4926 (Paper\,I)

\bibitem[\protect\citeauthoryear{Sebastian et al.}{2019}]{Sebastian2019}
Sebastian B., Kharb P., O'Dea C.~P., Gallimore J.~F., Baum S.~A., 2019, MNRAS, 490, L26

\bibitem[\protect\citeauthoryear{Swarup et al.}{1991}]{Swarup1991}
Swarup G., Ananthakrishnan S.,  Kapahi V.~K.,  Rao A.~P., Subrahmanya C.~R., Kulkarni V.~K., 1991, Curr. Sci., 60, 95

\bibitem[\protect\citeauthoryear{Wilkinson et al.}{1994}]{Wilkinson1994}
Wilkinson P.~N., Polatidis A.~G., Readhead A.~C.~S., Xu W., Pearson T.~J., 1994, ApJL, 432, L87

\end{thebibliography}
\end{document}